\documentstyle[here,epsfig]{mn}
\begin{document}
\def\simlt{\mathrel{\rlap{\lower 3pt\hbox{$\sim$}}\raise 2.0pt\hbox{$<$}}}
\def\simgt{\mathrel{\rlap{\lower 3pt\hbox{$\sim$}} \raise
2.0pt\hbox{$>$}}}

\title[The Radio Loud / Radio Quiet dichotomy: news from the 2dF QSO 
Redshift Survey ] 
{The Radio Loud / Radio Quiet dichotomy: news from the 2dF QSO Redshift Survey}
\author[M. Cirasuolo, M. Magliocchetti, A. Celotti, L. Danese] 
{M. Cirasuolo, M. Magliocchetti, A. Celotti, L. Danese  \\ 
SISSA, Via Beirut 4, 34014, Trieste, Italy }

\maketitle \vspace {7cm }

\begin{abstract}
We present a detailed analysis of a sample of radio-detected quasars,
obtained by matching together objects from the FIRST and 2dF Quasar 
Redshift Surveys. The dataset consists of 113 sources, spanning a redshift 
range $0.3 \simlt z \simlt 2.2$, with optical magnitudes 
$18.25 \le b_J \le 20.85$ and radio fluxes $S_{1.4{\rm GHz}} \ge 1$~mJy.
These objects exhibit properties such as redshift and colour distribution in 
full agreement with those derived for the whole quasar population, suggestive 
of an independence of the mechanism(s) controlling the birth and life-time 
of quasars of their level of radio emission.\\
The long debated question of radio-loud (RL)/radio-quiet (RQ) dichotomy is 
then investigated for the combined FIRST-2dF and FIRST-LBQS sample, since they 
present similar selection criteria. We find the fraction of radio detections 
to increase with magnitude from $\simlt 3$\% at the faintest levels up to 
$\sim 20$\% for the brightest sources.\\
The classical RL/RQ dichotomy, in which the distribution 
of radio-to-optical  ratios and/or radio luminosities shows a lack of
sources, is ruled out by our analysis.
We also find no tight relationship between optical and radio luminosities for 
sources in the considered sample, result that tends to exclude the mass of the 
quasar black hole as the physical quantity associated to the level of
radio emission.
\end{abstract}


\section{INTRODUCTION}

It was soon realized that not all the quasars, though first 
discovered by Schmidt in 1963 at
radio wavelengths, are powerful radio sources (Sandage 1965). Several optically
selected quasar samples have been observed in radio (e.g. Sramek \& Weedman 
1980; Condon et al. 1981; Marshall 1987; Miller, Peacock \& Mead 1990; 
Kellermann et al. 1989), showing that typically only 10\% - 40\% of 
the quasars are radio detected.

From these studies it was suggested that quasars
can be divided into the two different populations of ``Radio-Loud''  
and ``Radio-Quiet''  on the basis of their radio emission.
Kellermann et al. (1989), performing  
VLA observations of the Palomar-Green Bright Quasar Survey (the so-called PG 
sample), 
found that the radio-to-optical ratios of these objects -- defined as the 
ratio between radio and optical luminosity -- presented a bimodal 
distribution, clearly showing the occurrence of these two different 
populations.\\
Miller, Peacock \& Mead (1990) also found a dichotomy in the quasar population,
although this time based on radio luminosity as the parameter to define the 
level of radio loudness.

A step forward in the study of the radio properties of quasars came with the
FIRST Survey with the  VLA (Becker, White \& Helfand 1995) which was able to 
collect a large sample of quasars at faint flux levels.\\
Recent works based on this survey (FIRST Bright Quasar Survey; White et al.
2000 and Large Bright Quasar Survey; Hewett et al. 2001) 
suggest that the RL/RQ dichotomy could  be an effect 
due to the brighter radio and optical limits of the previous studies.
The issue is however still under debate. 
A recent work by Ivezic et al. (2002), 
based on the cross-correlation of the Sloan Digital Sky Survey with the FIRST 
Survey seems to find clear evidence for bimodality (see also Goldschmidt et 
al. 1999).

From the theoretical point of view,
despite the great advances in our ability of collecting unbiased
sets of data, the physical mechanism(s) responsible for the radio emission 
in Active Galactic Nuclei is still unclear. 
It is generally accepted to be related to the processes of 
accretion onto a central black hole (BH) --
the engine responsible for the optical-UV emission -- but 
no correlation  between radio and optical luminosity has been found so far
in such objects (see e.g. Stocke et al. 1992). On the other hand, 
although controversial, 
there is  some evidence for the fraction of radio-loud quasars to 
increase with
increasing optical luminosity (Padovani 1993; La Franca et al. 1994; Hooper 
et al. 1995; Goldschmidt et al. 1999; but see also Ivezic et al. 2002 for a 
dissenting view), and more recent findings seem to identify the angular 
momentum of a spinning BH -- extracted by a magnetic field -- as the correct 
mechanism to provide the necessary energy to fuel radio jets (for a rewiew see 
Blandford 2000) and therefore turn a radio-quiet object into a radio-loud one.

A different approach deals with the possibility for radio loudness to be 
connected with the intrinsic properties of the host galaxy. 
Early studies in fact concluded that, while radio-loud quasars
reside in elliptical hosts, radio-quiet ones are mainly found in spiral 
galaxies (Malkan 1984; Smith et al. 1986). Furthermore, it was observed a 
preference for radio-quiet quasars to be located in environments considerably 
less dense than those of radio-loud objects (Yee \& Green 1987; Ellingson 
et al. 1991). \\
More recent studies (Dunlop et al. 2002; Finn et al. 2001) however find 
a very different picture whereby the hosts of both radio-loud and radio-quiet 
quasars are massive elliptical galaxies with basic properties
(colors, environments, etc.) indistinguishable from those of quiescent, 
evolved, low-redshift ellipticals of comparable mass.

The aim of this 
work is then to analyze a wide sample of radio detected quasars drawn from 
the joined use of the FIRST and 2dF QSO Redshift Surveys in order to answer 
some of the questions raised throughout this section, with particular emphasis 
on the issue of radio-quiet/radio-loud dichotomy.

The layout of this paper is as follows. In Section 2 we give a brief 
description of the FIRST and 2dF datasets  and of the 
matching procedure used to cross-correlate them, while in Section 3 we analyse 
the properties of the sample obtained from the joined use of these two 
surveys. In Section 4 we study the dependence on redshift and optical 
luminosity of the fraction of radio-detected quasars and in Section 5 we 
discuss the problem of radio loudness with particular attention devoted to 
the issue of radio-loud/radio-quiet dichotomy. 
In Section 6 we summarize our conclusions. Throughout this
paper we will assume $H_0 = 50 \; {\rm km \; s^{-1} Mpc^{-1}}$, $q_0=0.5 $ and 
$\Lambda = 0$.   

\section{The Datasets}
\subsection{The FIRST Survey}

The FIRST (Faint Images of the Radio Sky at Twenty centimeters) survey (Becker
et al. 1995) began in 1993 and will eventually cover $\sim 10,000$
square degrees of the sky in the North Galactic cap and equatorial zones. The
beam-size at 1.4 GHz is 5.4 arcsec, with a rms sensitivity of typically 0.15
mJy/beam. Sources are detected using an elliptical Gaussian fitting 
procedure (White et al. 1997) with a $5 \sigma$ detection limit of 
$\sim 1$~mJy. The positional accuracy in the FIRST survey is $\simlt 0.5$arcsec 
at the 3 mJy level, reaching the value of 1 arcsec only at the survey threshold.

The latest release (5 July 2000) of the catalogue covers 7988 square degrees 
of the sky, including most of the area  $7^h20^m \simlt {\rm RA}({\rm 2000}) 
\simlt 
17^h20^m$, $22.2^\circ \simlt {\rm dec}({\rm 2000}) \simlt 57.5^\circ$ and 
$21^h20^m 
\simlt{\rm RA}({\rm 2000}) \simlt 3^h20^m$, $-2.8^\circ \simlt {\rm dec}
({\rm 2000}) 
\simlt 2.2^\circ$, and comprises approximately 722,354 sources down to a flux 
limit $S_{1.4 {\rm GHz}}\simeq 0.8$~mJy. The surface density of objects in the 
catalogue is $\sim 90$ per square degree, though this is reduced to $\sim 80$
per square degree if one combines multi-component sources (Magliocchetti et 
al. 1998). The survey has been estimated
to be 95 per cent complete at 2 mJy and 80 per cent complete at 1 mJy 
(Becker et al. 1995). Note that, as the completeness level quickly drops for flux 
levels fainter than 1 mJy, in the following analysis we will only consider sources 
brighter than this limit.
 
\subsection{The 2dF Quasar Redshift Survey}
For the purposes of this work we have considered the first release of the 
2dF QSO Redshift 
Survey, the so called {\it 2QZ 10k catalogue}. 
A complete description of the catalogue can
be found in Croom et al. (2001). Here we briefly recall its main properties. 
QSO candidates with $18.25 \le b_J \le 20.85$ were selected from the APM 
catalogue 
(Irwin, McMahon \& Maddox 1994) in two $75^{\circ} \times 5^{\circ}$ 
declination strips centered on 
$\delta=-30^\circ$ and $\delta=0^\circ$, with colour selection criteria 
$(u-b_j)\le 0.36$; $(u-b_j)< 0.12-0.8\;(b_j-r)$; $(b_j-r)< 0.05$.
Such a selection guarantees a large photometric completeness ($ > 90 $ \%)
for quasars within the redshift range $ 0.3 \le z \le 2.2$.

Redshifts for QSO candidates were determined via both cross-correlation of the 
spectra with specific templates (AUTOZ, Miller et al. in preparation) and by 
visual inspection. A flag was then assigned to each spectrum, where Q=1 
corresponds to high quality identifications and redshift determinations, Q=2 
means low-quality identifications and redshift determinations and Q=3 
indicates no redshift assignment.\\
Only 2dF fields with a spectroscopic completeness (defined as the ratio 
of objects observed in the field with Q=1 or Q=2 flags to the 
total number of spectroscopically observed objects) of 85 per cent or greater 
were included in this first release of the 2dF QSO catalogue. This corresponds 
to a mean overall completeness of 93 per cent which -- by also allowing for 
sources not yet observed in the targeted fields -- converts into an effective 
area for the survey of 289.6 square degrees (see Croom et al. 2001).

The final catalogue contains $\sim 21,000$ objects with reliable (Q=1; Q=2) 
spectral and redshift determinations, out of which $\sim 11,000$ are quasars 
($\sim 53$~\% of the sample).\\
Whenever available, the 2QZ 10k catalogue also includes radio fluxes at 
1.4 GHz from the NRAO VLA Sky Survey (NVSS; Condon et al. 1998) and X-ray 
fluxes from the ROSAT All Sky Survey (RASS; Voges et al. 1999).

\subsection{Matching Procedure}
The overlapping region between the FIRST and 2dF Quasar Redshift Surveys is 
confined to the equatorial plane: $ 9^h \; 50^m  \leq {\rm RA(2000)} \leq  14^h \;
50^m$ and $ -2.8^{\circ}  \leq {\rm dec(2000)} \leq 2.2^{\circ}$.\\ 
Optical counterparts for a subsample of FIRST radio sources have  
been obtained by matching together objects included in the radio catalogue 
with objects coming from the APM survey (Maddox et al. 1990) in the considered 
area (for a similar analysis see also McMahon et al. 2002).
By following this procedure, Magliocchetti \& Maddox (2002) find 4075 
identifications -- out of a total of $\sim 24,000$ $S_{1.4 {\rm GHz}}\ge 
1$~mJy radio sources -- in the APM catalogue down to $b_J\le 22$ and for 
a matching radius of 2 arcsec. This last value was chosen after a careful 
analysis as the best compromise to maximize the number of real associations 
(estimated to be $\sim 97$\%), and at the same time minimize the 
contribution from spurious identifications down to a negligible 5 per cent.

While this procedure has been proved to work for the population of radio 
sources as a whole since it mainly includes radio galaxies with point-like 
radio structures (Magliocchetti et al. 2002), caution needs to be taken when 
dealing with powerful radio sources such as quasars. These objects in fact 
often show multiple components such as jets and/or hot-spots.\\
The algorithm introduced by Magliocchetti et al. (1998) and adopted in this 
work to collapse sub-structured sources into single objects having radio 
fluxes equal to the sum of the fluxes of the various components, 
assigns positions to the final single-source products which correspond to 
the median point between the different sub-structures. 
Since multiple-component objects in general present quite complex morphologies,
it can be possible that a number of ``collapsed'' sources end up with radio 
positions for their centroids which are displaced from their optical 
counterpart by more than the originally chosen value of 2 arcsec.

For this reason, we relax the requirement on the matching radius to 5 arcsec 
and consider as 
true optical identifications all the radio-optical pairs offset by less than 
this figure.\\
Following the above procedure, we end up with 1044 identifications 
(hereafter called id-sample), down to a magnitude limit $b_J=20.85$, which 
optically show point-like structures typical of the population of QSOs 
(see Magliocchetti \& Maddox 2002 for further details).

Finally, in order to obtain redshift measurements and spectral features 
for these sources, we looked for objects in the 2QZ 10k catalogue with 
positions which differ by less than 2 arcsec from the optical positions of sources in 
the id-sample. The choice of this value for the matching radius is based on 
the 2 arcsec diameter of each 2dF fibre.\\
This procedure leads to 104 quasars included in the 2dF QSO catalogue and 
endowed with radio fluxes $S_{1.4 {\rm GHz}}\ge 1$~mJy. The distribution of 
offsets between optical and radio positions is shown in Figure 1.
Note that, despite the problems associated with multi-component structures 
discussed in this section, the overwhelming majority ($> 90$\%) of 
identifications still lie within 2 arcsec from the corresponding radio position. As 
expected, only a few sources (almost all associated to multi-component 
structures) present offsets $>2^{\prime\prime}$.

As a final remark we note that, even though a choice for a larger matching 
radius in principle increases the number of spurious associations, 
this is not a concern in our analysis given the low space-density of 2dF 
quasars. In fact, for a 5 arcsec matching radius, the expected number of 
random coincidences on an area of 122 square degrees (effective area of the overlapping 
region between the FIRST and 2dF QSO Surveys) is about $7\times 10^{-2}$. 
 
\begin{figure}
\center{{
\epsfig{figure=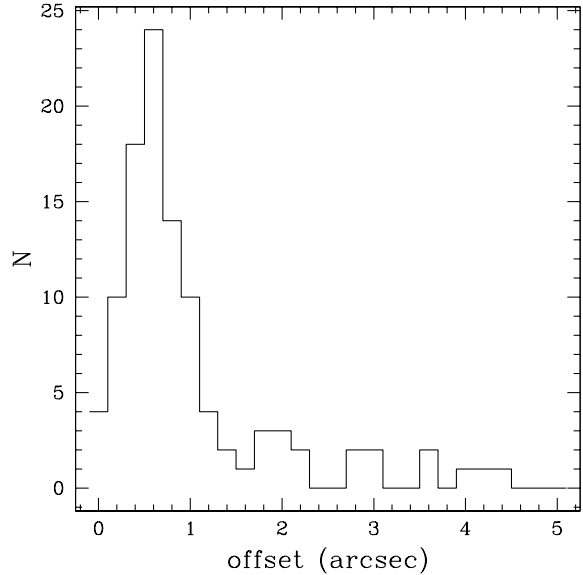,height=8cm}
}}
\caption{\label{offset} Distribution of offsets between radio and 
optical positions for the sub-sample of 2QZ 10k quasars included in the FIRST 
survey.}
\end{figure}

\subsection{FIRST vs NVSS}
To double check the
reliability of our sample and to also investigate the different efficiency of the
FIRST and NVSS surveys in detecting sources, we have then compared 
objects from the sample obtained as in section 2.3 with those sources 
included in the 
2QZ 10k catalogue and endowed with a radio-flux measurement from NVSS.

42 sources from the combined FIRST-2dF dataset do not present
NVSS flux measurements in the 2QZ catalogue. A direct search for these objects
in the NVSS on-line database found 14 of them in the 
flux range $3\simlt S_{1.4 {\rm GHz}}\simlt 600$~mJy. Out of these 14 sources,
three objects show multiple components and were therefore lost in the 
2dF-NVSS matching procedure due to lack of a combining algorithm 
for sub-structured objects. Three more are instead 
single sources which show quite large radio-to-optical offsets ($>7$~arcsec).
The others are single objects with offsets $\simlt 3$~arcsec and it is 
not clear why they were not included in the 2QZ 10k catalogue.\\
The remaining 28 sources included in the combined FIRST-2dF catalogue show fluxes 
$S_{1.4 {\rm GHz}}
\simlt 3$~mJy and were therefore lost by NVSS because of its relatively 
bright ($\sim 3$~mJy; Condon et al. 1998) flux limit.

Following the same procedure as above, we also found 9 of the 2dF quasars with 
NVSS fluxes to be missed by our matching procedure. After a direct search, 
it turned out that six of them are double/triple sources, with positions
for the centroids (as assigned by our combining procedure) which were displaced 
from the centers of the optical emission by more than 5~arcsec. The last 
three objects are instead point-like sources with radio-to-optical offsets 
between 5 and 7 arcsec and therefore lost by our searching criteria. These 
9 sources have been added to our original FIRST-2QZ 10k dataset and appear in
Table~1.

\begin{figure}
\center{{
\epsfig{figure=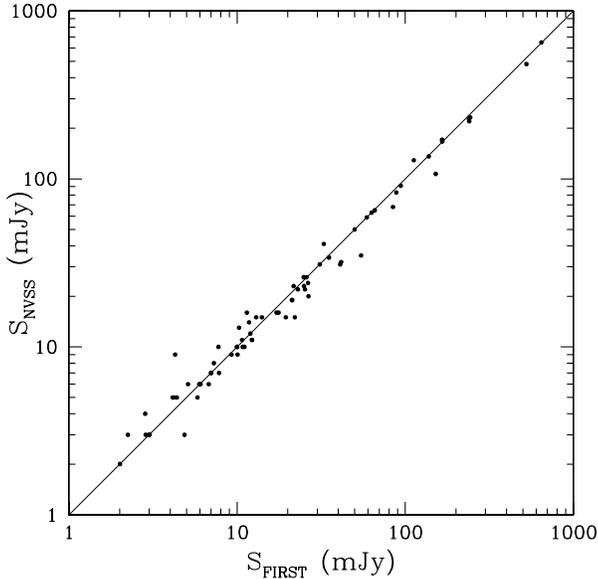,height=8cm}
}}
\caption{\label{diff_flux} Comparison between 1.4 GHz fluxes as measured by the 
FIRST and NVSS Surveys for sources in the FIRST-2dF sample.}
\end{figure}

As a final remark, note that the high resolution of the FIRST survey (beaming 
size 5.4~arcsec) might imply some of the flux coming from extended sources 
to be resolved out, leading to a systematic underestimate of the real flux 
density of such sources. In order to check for this effect, we have compared 
FIRST and NVSS (which, having a lower resolution -- beaming size of 45~arcsec 
-- should not be affected by this problem) fluxes for all the sources in the 
2QZ 10k catalogue which show a radio counterpart in both of these radio surveys. 
The result of the comparison is shown in 
Figure \ref{diff_flux}. It is clear that the agreement between 
fluxes as measured by FIRST and NVSS is excellent and  therefore no 
correction to the flux densities derived from the FIRST survey is needed.
%
%
\section{The Samples}

\subsection{FIRST-2dF}
Based on the procedure described in the previous section, the sample 
derived from the joined use of the FIRST and 2dF QSO surveys is 
constituted by 113 objects (hereafter called the FIRST-2dF sample) 
with optical magnitudes $18.25 \le b_J \le 20.85$ and radio fluxes
at 1.4 GHz $S_{1.4 {\rm GHz}}\ge 1$~mJy. All the objects included in the 
FIRST-2dF sample are presented in Table 1, whose columns respectively 
indicate: \\
(1) 2dF name; \\
(2) Right Ascension, RA(J2000) and Declination, dec(J2000) which -- 
except for objects with multiple sub-structures where the 
coordinates indicate the centroid of the source (obtained as in Magliocchetti et
al. 1998) -- correspond to the FIRST radio coordinates; \\
(3) Offset (expressed in arcsec) of the optical counterpart in the 
APM catalogue. An upper limit of 5 arcsec indicates those sources missed 
by our matching procedure, but found in the 2QZ 10k catalogue, as explained in 
section 2.4;  \\
(4) $b_J$ magnitudes; (5) $(u-b_J)$ and (6) $(b_J-r)$ colours; \\
(7) Radio-flux density (in mJy) at 1.4;\\
(8) Redshift; \\
(9) Notes on the radio morphological appearance, where {\it s} stands for 
point-like source, {\it d} for double source (i.e. presenting the 
two characteristic lobes) and {\it t} indicates a core+lobes structure. This 
information has been obtained by visual inspection of the images of each 
source from the FIRST atlas; \\
(10) The radio spectral index $\alpha_R$, whenever available (see later in this 
section).\\

\begin{figure}
\center{{
\epsfig{figure=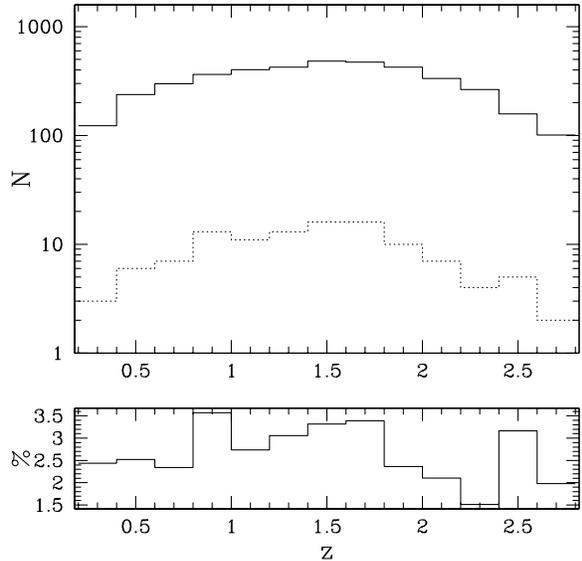,height=8cm}
}}
\caption{\label{zdist}Top panel: redshift distribution for the sample of 
FIRST-2dF QSOs (dotted line) and for all the quasars 
from the 2QZ 10k catalogue found in the North Galactic Cap (solid line).
Bottom panel: ratio between the two distributions.}
\end{figure}

The redshift distribution of these sources is shown in the top panel of 
Figure \ref{zdist} as a dotted line. 
For comparison, Figure \ref{zdist} also shows the redshift 
distribution of the $\sim 4,000$ quasars from the 2QZ 10k catalogue 
found in the North 
Galactic Cap (indicated by the solid line). Interestingly, 
the two distributions present the same trend, as they both smoothly rise 
for $z\simgt 0.3$, exhibit a maximum at about $z\sim 1.5$ and then decline 
at higher redshifts. This could be suggestive of a similar fuelling mechanism 
which controls the birth and life-time of quasars, regardless of their 
radio-emission.

Note that the observed trend for the two redshift 
distributions beyond $z\ge 2.1$ is biased by lack of completeness in the 
2QZ 10k sample due to colour selection effects. For this reason, 
in the following analysis we will only consider objects in the redshift range 
$0.35 \le z \le 2.1$.
The total number of sources in the FIRST-2dF sample is then reduced 
to 94, which also includes 3 Broad Absorption Lines QSO (BAL). 

The lower panel of Figure \ref{zdist} shows the
ratio between the two redshift distributions. This ratio is found to be 
$\sim 3$ \% over the entire redshift range. As it will be extensively 
discussed in the following sections, the joined effects of selecting sources 
in the blue band and of using a relatively faint magnitude range, makes the 
above figure lower than what previously found in literature ($\sim 10-20$ per 
cent; White et al. 2000; Hewett, Folz \& Chaffee 2001; Ivezic et al. 
2002).

\begin{figure}
\center{{
\epsfig{figure=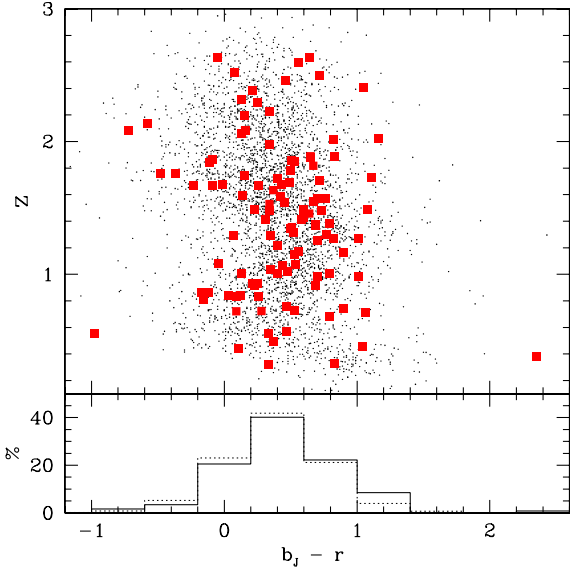,height=8cm}
}}
\caption{\label{br_z} Top panel: $b_J - r$ colours versus redshift for
the FIRST-2dF sample (filled squares) as compared with the 
ones obtained for all the 2QZ 10k quasars in the North Galactic Cap
(small dots). The lower panel represents the distributions of
$b_J - r$ colours for the FIRST-2dF sample (solid line) and for all the 2QZ 10k 
quasars, expressed as a percentage over the total number.}
\end{figure}

\begin{figure}
\center{{
\epsfig{figure=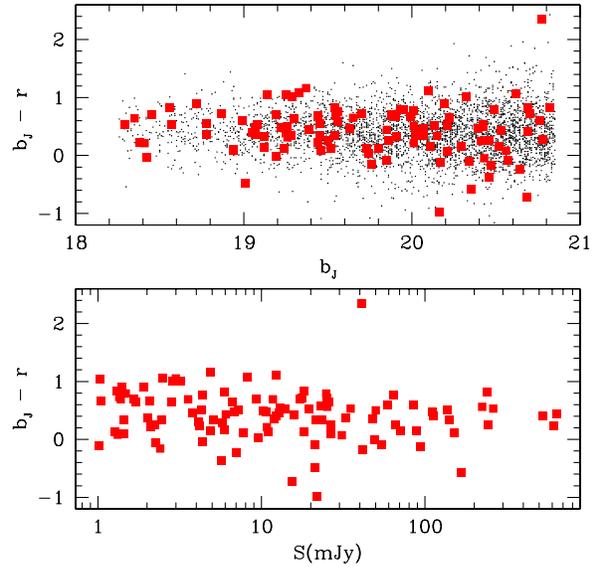,height=8cm}
}}
\caption{\label{br_s} $b_J - r$ colours versus $b_J$ magnitude (top
panel) and radio flux at 1.4 GHz (lower panel). Simbols are as in Figure
\ref{br_z}.}
\end{figure}

Figure \ref{br_z} shows the distribution of $b_J - r$ colours for the
sample of FIRST-2dF quasars (filled squares) as a function of redshift (top
panel); 
a large fraction of sources present values $0 \simlt  b_J - r \simlt 1$ (see
lower panel), independent of redshift. For comparison, 
the top panel in Fig. \ref{br_z} also shows, in small dots, the distribution of 
 $b_J - r$ colours obtained for all the 2QZ quasars included in the 
overlapping FIRST/2dF region (North Galactic Cap). As shown in Figure \ref{br_s}
the distribution of $b_J - r$ colours is also independent of apparent magnitude
and radio flux.
It is evident that radio-emitting sources follow 
the same distribution as the one obtained for the quasar population as a whole, 
suggestive of a radio activity not related to the colour of the 
source.\\
Note that two objects present colours which greatly differ from the average 
values, the first one showing $b_J -r > 2$ and the second one having 
$b_J -r \simlt -1$. This is presumably due to their faint optical magnitudes 
(found to be $\sim -21$ and $\sim -22$ respectively in the first and 
second case), which allow for a non-negligible light contribution from 
the host galaxy.

A further piece of information on the nature of sources in the FIRST-2dF sample 
is given by the investigation of their radio spectral index $\alpha_R$ 
\footnote{Throughout this work we define the radio flux density as 
$S_{\nu} \propto  \nu^{-\alpha_{R}}$}. 
In order to measure $\alpha_R$ we searched 
 for observations performed  at different (usually at 5 GHz) wavelengths
 using NED (NASA/Ipac Extragalactic Database). Unfortunately, as objects in 
our sample are relatively faint, we only managed to acquire radio fluxes at 
5 GHz for the ten most luminous  ($S_{1.4{\rm GHz}} \ge 100 $~mJy; see Table
1) ones. 
The distribution of their spectral indices is found to be completely uniform, 
with five steep spectrum ($\alpha_R > 0.5$) and four 
flat spectrum ($\alpha_R < 0.5$) sources, plus one object which presents a 
negative value for the radio spectral index ($\alpha_R =-0.6$).\\
As the remaining objects do not have any counterpart for radio
frequencies $\nu> 1.4$~GHz, this suggests that sources in our sample 
mainly present steep radio spectra, since values $\alpha< 0.5$ would make at least 
some of them observable in surveys performed e.g. at 5 GHz. 
With some confidence we can then assume 
most of the quasars in the FIRST-2dF sample to have a steep spectrum and 
associate them to a mean value for the radio spectral index, $\alpha_R=0.8$.

As a final step -- 
in order to facilitate comparisons between our 
results and those found in literature -- we decided to 
convert magnitudes from the $b_J$ to the B band. 
To compute the mean $B-b_J$ we used the composite quasar spectrum compiled 
by Brotherton et al. (2001) from $\sim 600$ radio-selected quasars in the 
FIRST Bright Quasar Survey (FBQS).
It turns out that in the redshift range $0.3 \le z \le 2.2$ 
the difference between corresponding values in the 
two bands is very small, $0.05 \simlt B-b_J \simlt 0.09$, independent of 
redshift. We therefore chose to apply a mean
correction $B= 0.07 + b_J$. 
Note that the k-correction in the B band has also been computed from the 
Brotherton et al. (2001) composite quasar spectrum.\\
%
\subsection{LBQS}
In the previous section we have shown the main observational properties of the
FIRST-2dF quasars. Even though $\sim 100$ radio-emitting QSOs represent one 
of the widest samples obtained so far, nevertheless --  in order to increase the 
statistics and cover larger portions of the $M_B - z$ plane -- 
we decided to combine our dataset with other existing samples. 
Thanks to its selection criteria, the Large Bright Quasar Survey (LBQS) then comes as 
the natural extension of our original dataset to
brighter magnitudes.\\
A detailed description of this survey  
can be found in Hewett et al. (1995). Briefly, it consists of quasars optically 
selected from the APM catalogue (Irwin et al. 1994) at bright ($b_J < 19$) 
apparent magnitudes. Redshift measurements were subsequently derived for 1055 
of them over an effective area of 483.8 square degrees. Due to the 
selection criteria of the survey, quasars were detected over a wide redshift 
range ($0.2 \le z \le 3.4$), with a degree of completeness  
estimated to be at the $\sim 90$\% level.

Recently, this sample has been cross-correlated with the FIRST radio survey 
(Hewett et al. 2001) by using a searching radius of 2.1 arcsec over 
an area of 270 square degrees. This procedure yielded a total 
of 77 quasars (hereafter called the FIRST-LBQS sample)
with radio fluxes  $S_{1.4 {\rm GHz}} \ge 1$~mJy, magnitudes in the range  
$16 \simlt b_J \simlt 19$, and fractional incompleteness of $\sim 10$\% .

\subsection{The Combined Sample}
\begin{figure}
\center{{
\epsfig{figure=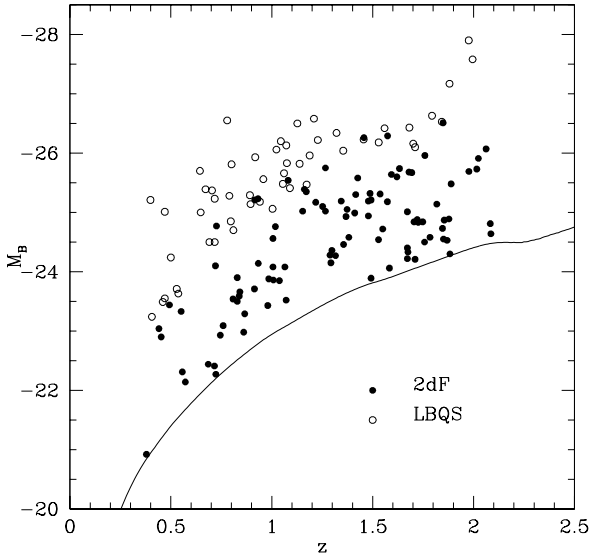,height=8cm}
}}
\caption{\label{mz} Absolute $M_B$ magnitude versus redshift for the 
FIRST-2dF (filled circles) and the FIRST-LBQS (empty circles) samples. 
The solid line describes the
selection effect due to the limiting magnitude $b_J=20.85$.}
\end{figure}

The joined FIRST-LBQS and FIRST-2dF samples span an extremely wide range in 
magnitudes -- $16.5 \simlt b_J \simlt 20.85$ -- providing a very good 
coverage of the $M_B - z$ plane. 
This can be seen in Figure \ref{mz} which represents the distribution of absolute 
$M_B$ magnitudes as a function of redshift for the two datasets.
In the redshift range where our sample 
is complete ($0.35 \le z \le 2.1$) a wide portion of the $M_B - z$ plane is 
filled by sources, and this obtained with the advantage of using 
only two samples,drawn from the same parent catalogues (FIRST and APM) and with 
compatible selection criteria.

\begin{figure}
\center{{
\epsfig{figure=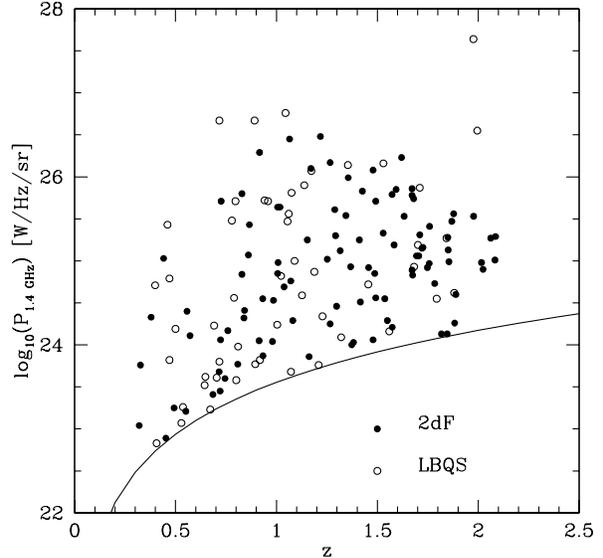,height=8cm}
}}
\caption{\label{pz} Radio-power $P_{1.4 {\rm GHz}}$ versus redshift for the FIRST-2dF
(filled circles) and the FIRST-LBQS (empty circles) samples.
The solid line describes the
selection effect due to the limiting flux density of 1 mJy.}
\end{figure}

We remind here that low-redshift, low-luminosity quasars with $M_B > -23$ 
have their optical emission dominated by the host galaxy, and will therefore 
be missed from the 2QZ catalogue as a result of the stellar appearance 
selection criterion applied to the input catalogue. 
Following the works by Boyle et al.(2000) and Croom et al.(2001), we then
exclude from the dataset all the low-luminosity quasars with $M_B > -23$.
As Figure 5 shows, this final sample exhibits a full 
coverage of the $M_B - z$ plane, except for the narrow region $-23\simlt M_B
\simlt -24$ at $z\simgt 1$.\\
Note that the radio power-redshift plane (see Figure \ref{pz}) is also widely covered;
only the region between $23\simlt {\rm log_{10} \: P_{1.4 \: GHz}}\:
({\rm W \: Hz^{-1} sr^{-1}})\simlt 24 $ is not entirely filled, 
due to the 1~mJy flux limit of the FIRST survey.

To summarize, our final sample is made of 141 quasars, either coming from 
the 2QZ 10k catalogue or from the LBQS dataset. These sources, brighter that $b_j=20.85$ 
and $M_B=-23$, are found in the redshift range $0.35 \le z \le 2.1$, and are endowed 
with a radio counterpart in the FIRST catalogue with $S_{1.4 {\rm GHz}}\ge 1$~mJy.
The sample is estimated to be $\simgt 90$\% complete with respect to the 
optical selection criteria and spectroscopic data acquisition, and 
80\% complete in its radio component (with a completeness level reaching 
100\% for fluxes brighter than 3~mJy; see Becker et al. 1995).\\
In the following sections we will then exploit the potentialities of this sample, 
homogeneous and deep both in radio and in optical, to investigate the 
properties of radio-active quasars. 
%
\section{Fraction of Radio Detected Quasars}

\begin{figure}
\center{{
\epsfig{figure=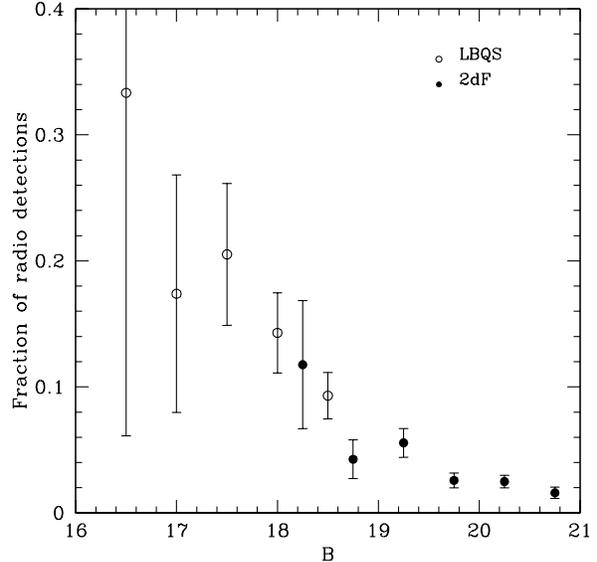,height=8cm}
}}
\caption{\label{RL_bj} Fraction of sources with a counterpart in the FIRST 
Survey ($S_{1.4 {\rm GHz}}\ge 1$~mJy) for the Large Bright 
Quasar Survey (empty circles) and 2QZ 10k (filled dots) sample as a function 
of apparent magnitude.}
\end{figure}

A key point on the properties of quasars is the determination of the fraction of
radio-emitting sources at a certain flux level, and its possible dependence
on redshift and optical luminosity.\\
As it was shown in Figure \ref{zdist}, from the FIRST-2dF sample we find 
this fraction to be relatively small ($\sim 3$\%) when compared to previous 
works ($\sim 10-20$ per cent; White et al. 2000; Hewett, Folz \& Chaffee 2001;
Ivezic et al. 2002), and independent of redshift.
The dependence of the fraction of radio-detected quasars on apparent B 
magnitude is instead presented in Figure \ref{RL_bj}, both for the FIRST-LBQS  
and the FIRST-2dF samples. Note that with ``radio-detected'' we indicate all the 
quasars having radio fluxes greater then 1~mJy and therefore observed by FIRST.\\
As already noticed by Hewett et al. (2001), we find a significant decrement 
for increasing magnitudes, with the fraction of radio-detections going from 
$\sim 20$\% at B=17 to $\sim 2$\% at B=21.
These values are very different from the 25\% of objects with magnitudes 
$E \le 17.8$ deduced by White et al. (2000) from the First Bright Quasar Survey
(FBQS). However, Hewett et al. (2001) point out that by applying corrections for the 
bandpass differences between the LBQS and the FBQS surveys (selected in $B_J$ and E band
respectively), the predicted fraction of
radio-detected quasars with magnitudes $E \le 17.8$ is found to be 15-17\%, 
which is close to the value obtained for the LBQS sample.

\begin{figure}
\center{{
\epsfig{figure=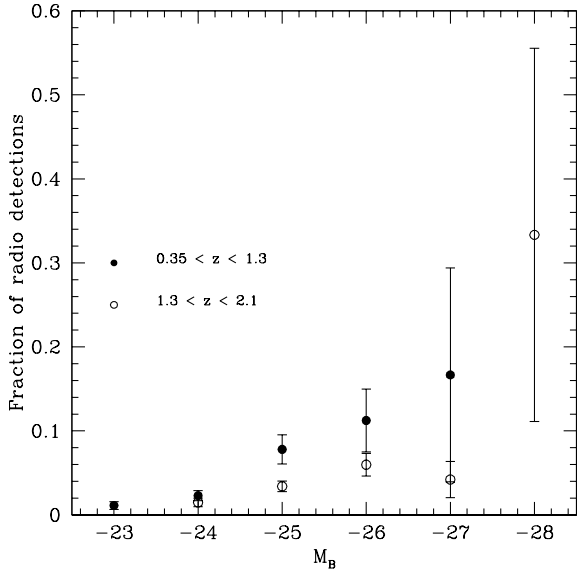,height=8cm}
}}
\caption{\label{RL_M} 
Fraction of sources with a counterpart in the FIRST 
Survey ($S_{1.4 {\rm GHz}}\ge 1$~mJy) from both the Large Bright 
Quasar Survey and 2dF QSO sample as a function of absolute magnitude $M_B$ 
in the two redshift ranges: $0.35 \le z \le 1.3$ (filled circles) and 
$1.3 < z \le 2.1$ (empty circles). Only sources with ${\rm log_{10} \: P_{1.4} \ge 
24 \;(W \; Hz^{-1} sr^{-1})}$ have been used in this case.}
\end{figure}

The observed behaviour of the fraction of radio-detected quasars 
could be explained as either due to an intrinsic dependence of this fraction on  
optical  luminosity, as already suggested by Miller et al. (1990), or as simply given 
by selection effects. In order to test these hypotheses, 
in Figure  \ref{RL_M} we have plotted the fraction of quasars from both the 
2QZ 10k sample and the LBQS that have a counterpart 
in the FIRST catalogue as a function of the
absolute magnitude $M_B$, and for the two different redshift bins 
$0.35 \le z \le 1.3$ and $1.3 < z \le 2.1$.\\
To minimize the selection effects, we have chosen to use only objects 
 with radio powers $\rm{log_{10} \: P_{1.4 \: GHz}} \ge 24 \;{\rm (W \; Hz^{-1} sr^{-1})}$,
since from Figure~\ref{pz} it can be seen that the range 
$23 \le {\rm log_{10} \: P_{1.4}(W \; Hz^{-1} sr^{-1})}<24$ is not entirely filled with 
sources due to the 1~mJy flux threshold of the FIRST. 

It turns out that the fraction of radio-detected quasars is indeed dependent on their 
optical luminosity: it grows from $\simlt 5$\%  in the case of  
faint ($M_B \sim -24$) objects, up to $\sim 20$\%  for the most powerful 
sources with $M_B \simlt -27$. This trend is present in both of the redshift bins, 
implying an  optical luminosity function for the radio-detected quasars 
flatter than the one measured for the quasar population as a whole.\\
This result is in agreement with what found by Goldschmidt
et al. (1999) from the Edinburgh Survey, who also observe the fraction of radio-detected 
QSOs to slightly decrease for increasing look-back times.
Our data show something similar but, 
because of the large error-bars associated to high optical luminosities, this 
finding does not have great statistical significance.\\
In fact, if we perform a Kolmogorov-Smirnov (KS) test we find that, while 
the hypothesis for the two samples of optical quasars and
radio-detected ones to have the same distribution in absolute magnitude  
is discarded at a very high significance level (probability $\sim 10^{-8}$), 
the same test  gives a probability of 
$\sim 0.7$ for the data sets to be drawn from the same redshift distribution.

%

\section{Radio Loudness}\label{radioloud}
%
In this section we will tackle in more detail the issue of radio loudness, 
with particular attention devoted to the problem of  
RL/RQ dichotomy.\\
Two parameters have been proposed to define the radio-loudness of a quasar: the 
first one is the radio-to-optical ratio $R^*_\nu$ defined as the ratio 
between the rest frame radio luminosity at a given frequency $\nu$ and the 
optical luminosity usually in the B band (Kellermann et al. 1989). 
Miller et al. (1990) however argued that $R^*_\nu$ has a physical 
meaning only if radio and optical luminosities are
linearly correlated, and therefore choose the radio power $\rm P_\nu$ as a better 
parameter to describe the level of radio loudness.

Several studies which used radio surveys at 5 and 8.5 GHz  
(Stocke et al. 1992; Padovani 1993; Hooper et al. 1995; Goldschmidt et
al. 1999) argued for a gap in the 
distribution of radio powers and/or radio-to-optical ratios for the 
objects under exam. The same behaviour was also recently claimed by 
Ivezic et al. (2002). The presence of
a bimodal distribution has been interpreted in the past as direct evidence for  
quasars to be divided into the two distinct populations of
 ``radio-loud '' and ``radio-quiet'',  with different properties and 
probably also different mechanisms responsible for the radio emission.
The threshold values at which the radio-quiet/radio-loud transition would 
happen have been inferred by these early works to be $R^*_{8.5 \:{\rm GHz}}\sim 
10$ (corresponding to $R^*_{1.4 \:{\rm GHz}}\sim 30-40$ for objects with 
spectral index $\alpha_R=0.8$) or ${\rm log_{10} \: P_{5 GHz} \sim 24 
\;\; (W \; Hz^{-1} sr^{-1})}$.
\begin{figure}
\center{{
\epsfig{figure=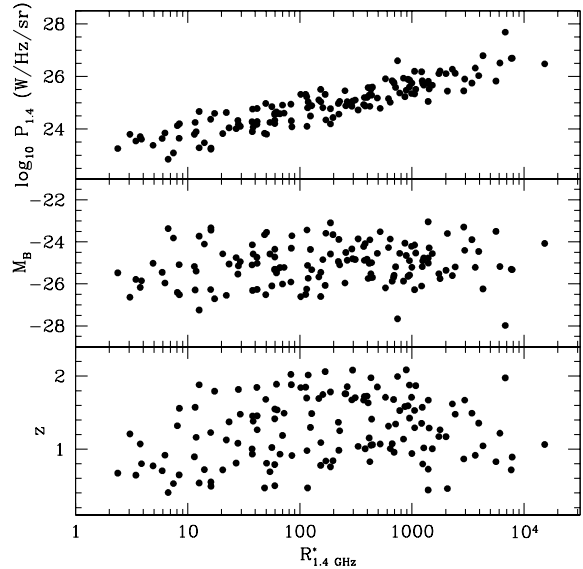,height=8cm}
}}
\caption{\label{r_p_z} Radio-to-optical radio $R^*_{1.4 {\rm GHz}}$ versus radio power
(top panel), absolute magnitude (middle panel) and redshift (bottom panel) for
the sources in our combined (FIRST-2dF and FIRST-LBQS) sample.}
\end{figure}

It is worth remarking here that the above definitions are not equivalent. In fact, 
if we consider our combined FIRST-2dF and FIRST-LBQS dataset and plot the 
radio-to-optical ratio $R^*_{1.4 {\rm GHz}}$ as a function of radio luminosity 
${\rm P_{1.4\:GHz}}$ (top panel of Figure~\ref{r_p_z}), it becomes clear 
that not all the sources (although still a great portion of them) satisfying 
one of the two criteria can be accepted as radio-loud by the other one.
This is the reason why in the following sections -- in order to assess the 
possible presence of a RL/RQ dichotomy in our sample --  
we will consider both the $R^*_{1.4 {\rm GHz}}$ and ${\rm P_{1.4\:GHz}}$ distributions.\\
Note that if we apply the above definitions for radio loudness to the combined sample, 
we then find $\sim 75$\% of the sources to be considered as ``radio-loud'', with the  
remaining $\sim 30 - 40$ objects probing part of the ``radio quiet'' regime. 

As a last consideration before investigating in more detail the issue of 
dichotomy, we note that the distribution of 
radio-to-optical ratios for the combined sample is totally independent of 
both optical luminosity (middle panel of Figure~\ref{r_p_z}) and 
redshift (bottom panel of Figure~\ref{r_p_z}). This last finding is in 
agreement with the results obtained by White et al. (2000) for the First 
Bright  Quasar Survey and Hewett et al. (2001) for the Large
Bright Quasar Survey.
\begin{figure}
\center{{
\epsfig{figure=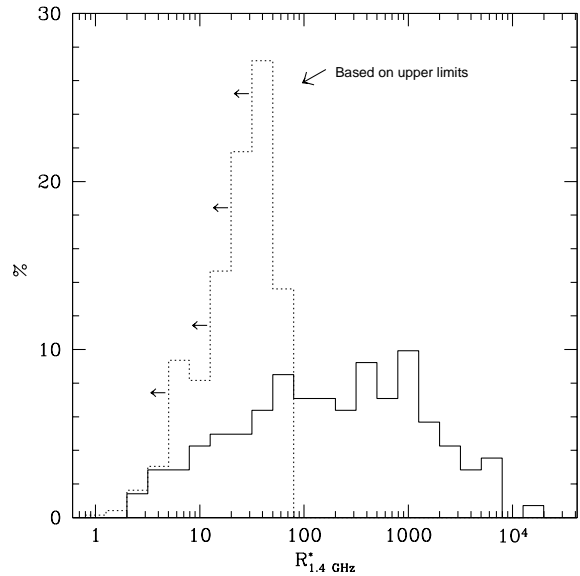,height=8cm}
}}
\caption{\label{historo} Distribution of radio-to-optical ratios $R^*_{1.4 {\rm GHz}}$ 
for the sources in the combined sample expressed in percentages (solid line). 
The dotted histogram has been obtained for all the quasars found in the 
North Galactic Cap of the 2QZ 10k 
catalogue and the Large Bright Quasar Sample which do not have a radio 
counterpart in the FIRST survey, by assuming an upper flux limit of 1~mJy 
(see text for details).}  
\end{figure}
%
\subsection{No Evidence for a Bimodal Distribution}
\begin{figure}
\center{{
\epsfig{figure=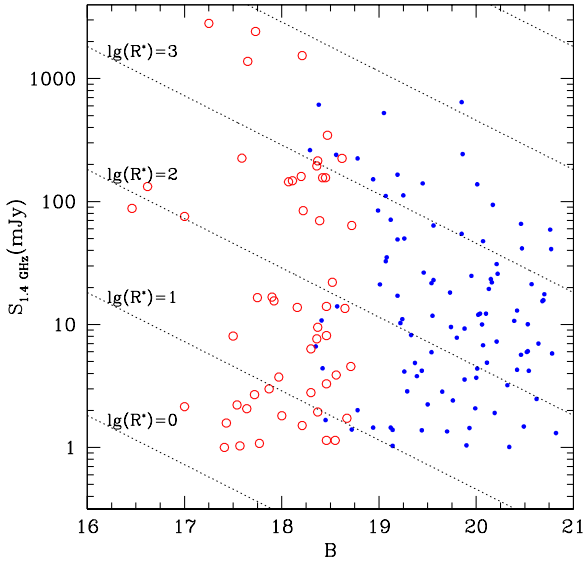,height=8cm}
}}
\caption{\label{flux_bj} Radio flux S versus B magnitude for the FIRST-2dF 
(filled dots) and FIRST-LBQS (empty circles) samples. The dotted lines
are the loci of constant radio-to-optical ratio $R^*_{1.4 \:{\rm GHz}}$ 
(see text for details).}
\end{figure}

The previous definitions of radio loudness were introduced in literature because 
the distributions of radio power and $R^*$ for the objects under exam appeared to be
bimodal.
In order to test the presence of this bimodal behaviour in our dataset,
we have then considered the distribution of radio-to-optical ratios 
$R^*_{1.4\:{\rm GHz}}$  for objects in the
combined sample. As Figure~\ref{historo} illustrates, our data (shown by the solid 
line) do not present any gap around $R^*_{1.4 {\rm GHz}} \sim 40$, and the distribution 
appears to be quite flat over a large range of radio-to-optical ratios, 
the decline observed for $R^*_{1.4 {\rm GHz}} \simlt 30$ being fully consistent with
effect of the  1~mJy flux limit of the FIRST survey. \\
This effect becomes more clear if one considers Figure~\ref{flux_bj}, where
the apparent optical magnitude $B$
has been plotted as a function of the radio flux at 1.4 GHz both for the FIRST-2dF 
and the FIRST-LBQS samples.
The dotted lines represent the loci of constant radio-to-optical ratio. 
One can then see that the 1~mJy radio limiting flux 
determines a loss of $R^*_{1.4 {\rm GHz}} \simlt 30$ sources 
which becomes progressively more relevant as the value of the 
radio-to-optical ratio associated to such objects decreases.

To have an idea of the $R^*$ distribution of these ``missing'' sources, we 
can plot (as shown by the dotted line in Figure~\ref{historo}) the distribution of 
radio-to-optical ratios for those quasars found in the North Galactic Cap 
of the 2QZ 10k Survey and in the Large Bright Quasar Survey which do not have a
counterpart in the FIRST catalogue, where 
values for $R^*$ have been calculated by assuming an upper flux limit of 1~mJy.
We can reasonably conclude that 
our data show no evidence for a $R^*_{1.4 {\rm GHz}} \sim 30-40$ gap, 
in agreement e.g. with the White et al.(2000) results on the FBQS sample.\\
The presence of a bimodal distribution in 
the population of quasars has been recently re-claimed by Ivezic et al.(2002) 
who analysed a sample of 
radio sources coming from the joined use of the FIRST and Sloan 
Surveys. We believe their results to be highly biased by the cut 
both in flux and in magnitude, which preferentially excludes sources with low 
values of the radio-to-optical ratio and therefore generates an artificial 
gap in the distribution of $R^*$ for 
${\rm log_{10}}R^*_{1.4 {\rm GHz}}\simlt 2.5$. 
In fact, the authors select objects in a region of the apparent 
magnitude-radio 
flux plane enclosed by lines perpendicular to the loci of constant radio-to-optical
ratios (see their Figure 14). However, we argue that in order to obtain an unbiased
$R^*_{1.4}$ distribution, a weight should be applied to take into account
optical number counts, 
since their steep rise for increasing magnitudes
determines a spurious detection of both the apparent peak at large values of $R^*_{1.4}$ 
and a lack of sources with low radio-to-optical ratios.

\begin{figure}
\center{{
\epsfig{figure=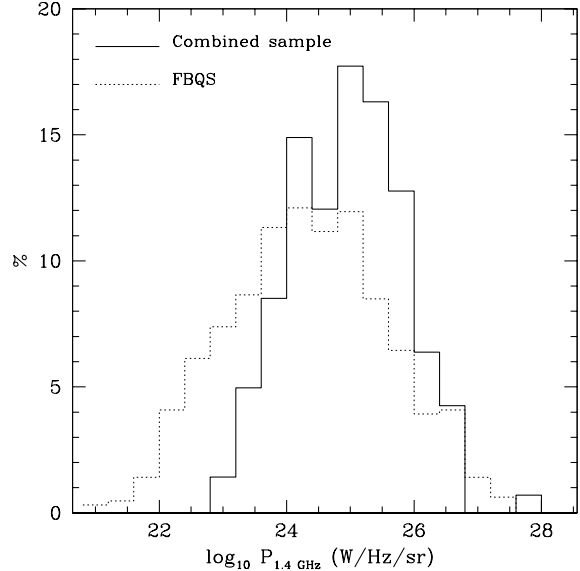,height=8cm}
}}
\caption{\label{histo_pradio} Distribution of radio powers 
for sources of the combined FIRST-2dF and FIRST-LBQS sample (solid line) 
and in the case of objects from the FBQS (dotted line).}
\end{figure}

No evidence for a ``gap'' is seen even 
if we use the radio-power $\rm P_{1.4 GHz}$ as the parameter which separates 
quasars into 
radio-quiet and radio-loud ones.
 We remind here that the assumption for a bimodal 
distribution as expressed above would in fact imply a deficit 
of sources at radio luminosities ${\rm log_{10} \: P_{1.4 GHz} \sim 24.5 \;\;
(W \; Hz^{-1} sr^{-1}})$ (assuming a spectral index  $\alpha_R = 0.8$). 
The absence of a $\rm P_{1.4 GHz}\sim 24.5$ gap is clearly shown 
in Figure~\ref{histo_pradio}.
Again, the decline observed 
for ${\rm log_{10} \: P_{1.4} \simlt 24 \;\; (W \; Hz^{-1} sr^{-1})}$ is due 
to the lack of low-power sources in our sample caused by the 1~mJy 
completeness limit of the FIRST survey (see Fig.~\ref{pz}).
Note that a very  similar trend is obtained for 
quasars from the FBQS (dotted line in Figure~\ref{histo_pradio}). 
Because of the broader redshift range spanned by this survey
(different selection criteria allow to also detect low -- $z\le 0.35$ -- 
redshift quasars), the FBQS can provide a better
coverage of the faint end of the radio power distribution, even though the 
1~mJy limit once again determines the decline of the distribution at low -- 
${\rm log_{10} \: P_{1.4 GHz} \le 23.5 \;\;(W \; Hz^{-1} sr^{-1}})$ -- radio powers. 
It is nevertheless clear that none of the two distributions hints to the 
presence of a gap for ${\rm log_{10} \: P_{1.4 GHz} \sim 24.5 \;\;(W \; Hz^{-1} 
sr^{-1}})$, i.e. to a bimodal distribution for the quasar population.
\begin{figure}
\center{{
\epsfig{figure=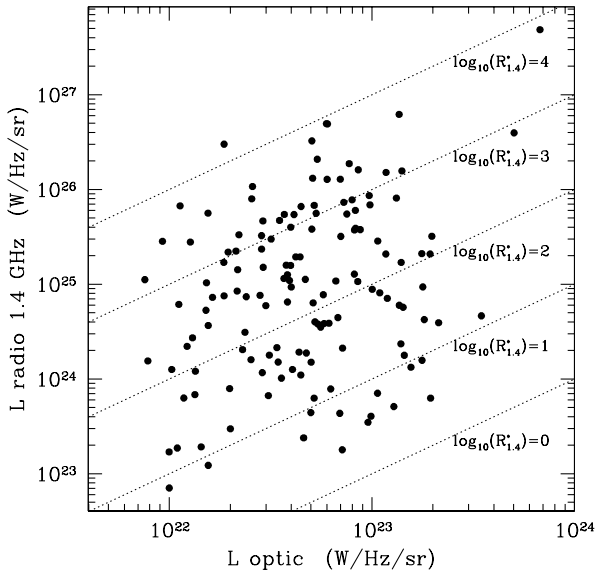,height=8cm}
}}
\caption{\label{lolr} Optical luminosity versus radio power for sources in
the combined FIRST-2dF and FIRST-LBQS sample. The dotted lines are the loci of 
constant radio-to-optical ratio. }
\end{figure}

An alternative way to look at this result is to consider the distribution 
of radio powers for the objects in our sample as a 
function of their optical luminosities.
As Figure~\ref{lolr} shows, the coverage of the L$_{{\rm radio}}$-
L$_{{\rm optic}}$ plane is homogeneous, and the transition from 
those objects defined as ``radio-loud'' to those belonging to the ``radio-quiet'' 
population extremely smooth. Again, no gap is present, implying   
a continuous variation in the radio properties of quasars.
This is in agreement with e.g. what found by Lacy et al. (2001) from the 
analysis of sources in the FBQS.

As a final remark, we note (see Figure~\ref{lolr}) that there is no tight relationship 
between radio power and optical luminosity for the objects in our combined 
sample, since sources with a 
particular luminosity L$_{{\rm optic}}$ can be endowed with radio powers 
spanning up to three orders of magnitude. 
From the above result we can also have some hints that the radio properties of quasars 
are not related to the mass of the central blach hole.
In fact, in the case of quasars it has been shown that the optical emission 
is tightly related to the  bolometric one (Elvis et al. 1994). 
We can  reasonably assume that these bright quasars
are emitting close to the Eddington limit and in this case the  bolometric luminosity
can in turn  be connected to the mass of the central black hole. Since we find a very
spread relation between radio and optical luminosity, it seems unlikely that the mass of 
the central black hole  dominates the level of radio emission.
%
\section{Conclusions}
We have presented a new sample of radio-emitting quasars, obtained by matching
together objects from the FIRST and 2dF Quasar Redshift Surveys. The dataset includes 
113 quasars, found within the redshift range $0.3 \leq z \leq 2.2$, with optical 
magnitudes $18.25 \leq b_J \leq 20.85$ and radio fluxes at 1.4 GHz $S \geq 1$~mJy.\\
This sample has then been combined with the FIRST-LBQS catalogue (Hewett et al. 2001) 
in order to provide an 
almost-complete coverage of the optical luminosity-redshift plane and increase the 
statistical significance of our results. The main conclusions deriving from the joined 
analysis of the two samples can be summarized as follows:

\begin{enumerate} 
\item The properties of radio-emitting quasars, such as their redshift
distribution and $ b_j  - r $ colours, are in full agreement with those derived for the
quasar population as a whole. This suggests the fuelling mechanism(s) responsible for 
the birth, colour and life-time of quasars to be independent of
the level of radio emission.  
\item The fraction of radio detections decreases for fainter apparent magnitudes.
This is also true if one considers the intrinsic luminosity of the sources, as the 
fraction of radio detections is found to grow from $\simlt 3$\% at $M_B\sim -24$ up to 
20-30 \% for the brightest ($M_B \sim -28$) objects.
\item The classical radio loud/radio quiet dichotomy, in which the distribution 
of radio-to-optical  ratios and/or radio luminosities shows a ``gap'',
 has been ruled out by our analysis. We found no lack of sources neither 
for $R^*_{1.4 {\rm GHz}}\sim 30-40$ nor for $ \rm {log_{10} \: P_{1.4} \sim 24 \;\; 
(W \; Hz^{-1} sr^{-1})}$.
Due to the selection effects -- in particular the 1~mJy radio flux limit of the FIRST 
survey -- we could not explore the low-P/low-$R^*$ regions of the distributions. 
However, at least in the observed ranges, it is  
suggestive of a smooth transition from the radio-loud to the radio-quiet regime. 
\item We find no tight relationship between radio and optical luminosity for the sources
in our sample. For a given optical luminosity the scatter in the radio power is
found to be more than three orders of magnitude. This result can give us some 
insight on the physical quantities responsible for the radio emission.
In fact -- since in the case of quasars the optical emission is tightly related to the
bolometric one (Elvis et al. 1994), the bolometric luminosity ( under the assumption that
quasars emit close to the Eddington limit) can in turn 
be connected to the mass of the central black hole -- such a large scatter 
tends to exclude the mass of the central black hole as the relevant quantity which 
controls the level of radio activity.

\end{enumerate}

It is clear that the above results -- especially those concerning the 
radio-quiet/radio-loud dichotomy -- need further investigation. While awaiting for 
wide-area surveys able to probe the low radio-power/low radio-to-optical ratio regime, 
we are testing the nature and behaviour of the radio emission of quasars,
through Monte Carlo simulations. We plan to tackle this 
issue in a future paper.
%
\section*{Acknowledgments}
We acknowledge the Italian MIUR and ASI for
financial support.
%

\end{document}